\documentclass[12pt]{article}
\usepackage[utf8]{inputenc}
\usepackage[T1]{fontenc}
\usepackage{mathptmx}
\usepackage[left=2.5cm, right=2.5cm, top=2.5cm, bottom=2.5cm]{geometry}
\usepackage{changepage}
\usepackage{hyperref}
\usepackage{booktabs}
\usepackage{caption}
\usepackage[backend=biber, style=apa, natbib=true]{biblatex}
\usepackage{epsfig,amsfonts,amsmath,graphicx,bm,url,float}
\usepackage{mathtools,eucal}
\usepackage{xr}
\usepackage{color}
\usepackage{subfig}
\usepackage{multirow}
\usepackage{tikz}
\usepackage{soul}
\soulregister{\texttt}{1}
\usetikzlibrary{shapes.geometric, arrows, shadows, fadings}

\title{\textbf{Efficient Bayesian inference for non-linear association structures in joint models: A hierarchical approach via INLA}}

\author{\textbf{Denis Rustand$^{1}$}, \textbf{H{\aa}vard Rue$^2$}, \textbf{Lisa Le Gall$^{1}$} and \textbf{Karen Leffondre$^{1}$}\\ \ \\
\small $^1$Bordeaux Population Health Research Center, Inserm U1219, University of Bordeaux, Bordeaux, France. \hfill\\
\small $^2$Statistics Program, Computer, Electrical and Mathematical Sciences and Engineering Division, \hfill\\
\small King Abdullah University of Science and Technology (KAUST),\\
\small Thuwal 23955-6900, Kingdom of Saudi Arabia\\}

\date{}

\begin{document}

\maketitle

\begin{abstract}
\begin{adjustwidth}{40pt}{40pt}
\normalsize
Joint models for longitudinal and time-to-event data are increasingly used in health research to characterize the association between biomarker trajectories and the risk of clinical events. However, these models usually assume a linear relationship between the longitudinal marker and the log-hazard of the event. This assumption is rarely verified and often fails to capture complex biological mechanisms, such as U-shaped risk profiles or plateau effects. In this paper, we propose a fast and stable hierarchical framework for non-linear association structures in joint models using Integrated Nested Laplace Approximations (INLA), implemented in the \texttt{INLAjoint} R package. Our approach builds upon a unified framework where the scaling effect of the marker is decomposed into a parametric baseline (constant and linear components) and a data-driven smooth deviation modeled via an orthogonal basis derived from a second-order random walk. This natural hierarchy allows researchers to adapt model flexibility directly and verify the linearity assumption using standard information criteria. Through simulation studies, we demonstrate that the proposed method accurately recovers complex non-linear trajectories. We illustrate the practical utility of our framework by analyzing the joint association of the current value and current slope of body mass index (BMI) with all-cause mortality in the Health and Retirement Study. This analysis reveals a U-shaped mortality risk for the BMI value, and a non-linear effect for the rate of weight change, where a declining weight trajectory is associated with higher mortality risk. 
\end{adjustwidth}
\end{abstract}

\normalsize
\section{Introduction} \label{sec:intro}

In biomedical, clinical and epidemiological research, studies generate longitudinal biomarkers measurements and time-to-event data. Historically, these processes were analyzed separately. Longitudinal data were usually evaluated using mixed-effects models to characterize population-average trends and subject-specific trajectories over time. Time-to-event outcomes were analyzed using standard survival models to estimate the relative risk of an event associated with baseline or time-varying covariates \citep{therneau2000cox}.

Analyzing these processes separately introduces methodological limitations when the longitudinal biomarker is endogenous \citep{wulfsohn1997joint}. Because the marker's trajectory is linked to the event status, it is subject to informative censoring. Moreover, treating noisy, intermittently observed endogenous markers as external, error-free covariates in standard time-dependent Cox models \citep{prentice1982covariate} causes regression dilution bias, attenuating association estimates toward the null \citep{sweeting2011joint}.

Several methods have been used to bypass the complexity of full joint likelihood maximization. The two-stage approach decomposes the estimation process into two sequential steps by first fitting a mixed-effects model and then inserting these predicted biomarker values into a time-dependent Cox model \citep{ye2008semiparametric}. While mitigating measurement error, this approach ignores the survival data in the first stage and fails to account for informative censoring.  Moreover, treating predicted trajectories as observed covariates (i.e., without error) fails to propagate estimation uncertainty from the longitudinal submodel to the survival submodel.

To avoid these biases, the joint modeling framework was developed \citep{wulfsohn1997joint, tsiatis2004joint}. Joint models simultaneously evaluate the longitudinal and time-to-event processes by linking them through a shared latent structure. By defining a joint likelihood that marginalizes over the random effects, joint models account for biomarker measurement error, handle irregularly spaced observation times and correct for informative dropout induced by the survival event \citep{rizopoulos2012joint}.

Despite these theoretical advantages, the vast majority of joint modeling applications rely on a highly restrictive structural assumption. They assume that the association between the longitudinal marker and the log-hazard of the event is linear. Mathematically, this implies that every one-unit increase in the shared latent component multiplies the hazard by a constant factor. Human physiology, however, frequently exhibits complex, non-linear patterns characterized by thresholds, plateau effects and U-shaped curves. For example, in a recent nephrology study investigating the longitudinal trajectory of serum uric acid in patients with chronic kidney disease, \citet{prezelin2023longitudinal} demonstrated a complex association where the hazard of kidney failure plateaus between specific levels before increasing exponentially beyond a threshold. In the same study, the authors found that the hazard of death before kidney failure exhibited a distinct U-shaped curve. Forcing a single linear parameter to summarize this U-shaped risk curve averages the opposing risk directions of the tails, leading to a near-zero coefficient that falsely implies no significant association. To capture these non-linear dynamics, \citet{prezelin2023longitudinal} used the biased two-stage approach. In their discussion, they highlighted this methodological compromise, noting that while a joint analysis would have been optimal, available software was lacking. This lack of accessible software forces researchers to either make unrealistic statistical assumptions or discard longitudinal data entirely. Illustrating the latter, \citet{sun2025non} investigated a non-linear threshold effect in HIV research, demonstrating that mortality risk drops as CD4+ T-cell counts increase until flattening at a protective level. However, to model this non-linear relationship, the authors relied on a single measurement of CD4+ T-cell counts per individual, ignoring the longitudinal nature of this marker.

Relaxing the linear assumption to accommodate non-linear association structures has become an important frontier in biostatistical methodology. \citet{kohler2018nonlinear} introduced a flexible Bayesian framework using P-splines to capture non-linear association structures in a joint model. However, their approach relies on Markov Chain Monte Carlo (MCMC) sampling, which imposes a significant computational burden. In their simulation studies evaluating a standard non-linear association, model estimation required an average of 3.9 hours for a cohort of 300 subjects and 7.2 hours for 600 subjects. Moreover, MCMC estimation for these highly parameterized joint models with non-linear associations frequently suffers from numerical instability due to poor chain mixing. The authors reported initial convergence failures in 15.5\% of simulated datasets for 300 subjects, requiring manual algorithm restarts with different random seeds to achieve stable estimates. These limitations prevent the routine adoption of such joint models with non-linear associations in applied research, leaving biostatisticians without practical tools to diagnose or model deviations from linearity.

To address this gap, we propose a fast and stable hierarchical framework to estimate non-linear association structures using Integrated Nested Laplace Approximations (INLA) \citep{rue2009approximate, vanniekerk2023new}. Unlike sampling-based methods, INLA uses deterministic approximations and sparse matrix algebra. This offers substantial improvements in stability and computation time for joint models \citep{rustand2023bayesian, rustand2024fast, rustand2026bayesian}. We introduce a unified hierarchical strategy for modeling associations that allows researchers to adapt model complexity directly to the data. This framework parameterizes the scaling effect of the marker into a constant baseline, a linear trend and a flexible data-driven deviation term based on the orthogonal basis expansion of a second-order random walk. This approach enables the estimation of complex risk trajectories in seconds rather than hours without restrictive parametric assumptions, making sensitivity analysis of the linearity assumption a feasible routine step in biostatistical workflows. To ensure wide accessibility and replicability, our proposed methodology is fully implemented in the free and open-source \texttt{INLAjoint} R package \citep{rustand2026bayesian}, and we provide reproducible code based on a simulated dataset.

The remainder of this paper is organized as follows. Section \ref{sec:methods} details the formulation of the joint models with non-linear associations within the INLA framework. Section \ref{sec:simulations} evaluates the frequentist properties and robustness of the method through simulation studies. Section \ref{sec:application} demonstrates the practical utility of this method by analyzing the relationship between longitudinal BMI dynamics (both current value and current slope) and mortality in the Health and Retirement Study (HRS). We conclude with a discussion in Section \ref{sec:discussion}.

\section{Methodology} \label{sec:methods}

\subsection{Joint model specification}
Let $\mathcal{D}_N = \{(\bm{y}_i, T_i, \delta_i), i=1, \dots, N\}$ denote the observed data for $N$ subjects. For each subject $i$, we observe longitudinal measurements $\bm{y}_i = (y_{i1}, \dots, y_{in_i})^\top$ at discrete time points $\bm{t}_i = (t_{i1}, \dots, t_{in_i})^\top$ and an event time $T_i = \min(T^*_i, C_i)$ with event indicator $\delta_i = \mathbb{I}(T^*_i \le C_i)$, where $T^*_i$ is the true event time and $C_i$ is the censoring time.

The longitudinal marker is assumed to follow a generalized linear mixed model framework. We define the observed outcome $y_{ij}$ as a noisy realization of a continuous-time latent process $\eta_i(t)$. For a continuous marker with Gaussian error, the submodel is specified as:
\begin{align} \label{eq:long}
    y_{ij} &= \eta_i(t_{ij}) + \varepsilon_{ij}, \quad \varepsilon_{ij} \sim \mathcal{N}(0, \sigma_e^2), \nonumber \\
    \eta_i(t) &= \bm{x}_i^\top(t) \bm{\beta} + \bm{z}_i^\top(t) \bm{b}_i,
\end{align}
where $\bm{\beta}$ denotes the fixed effects associated with covariates $\bm{x}_i(t)$ and $\bm{b}_i \sim \mathcal{N}(\bm{0}, \bm{\Sigma}_b)$ represents the subject-specific random effects associated with $\bm{z}_i(t)$. This defines the true unobserved trajectory $\eta_i(t)$. While we present the Gaussian case for simplicity, the framework extends naturally to non-Gaussian markers via the generalized linear mixed model framework.

The survival process is defined by the hazard function $\lambda_i(t)$, which depends on the longitudinal process through a shared latent component $\nu_i(t)$:
\begin{equation} \label{eq:hazard}
    \lambda_i(t) = \lambda_0(t) \exp\left( \bm{w}_i^\top(t) \bm{\varphi} + f(\nu_i(t)) \right).
\end{equation}
Here, $\lambda_0(t)$ is the baseline hazard, $\bm{w}_i(t)$ represents the vector of survival covariates and $\bm{\varphi}$ denotes the corresponding vector of survival fixed effects. The shared component $\nu_i(t)$ can take various functional forms. Common implementations include the current value ($\nu_i(t) = \eta_i(t)$) and the current slope ($\nu_i(t) = \frac{\partial}{\partial t}\eta_i(t)$). For notation, we write $\nu$ for a specific value of $\nu_i(t)$. In the standard joint model, the contribution to the log-hazard is linear, with $f(\nu)=\gamma\nu$, where $\gamma$ is a constant scaling association parameter.

\subsection{Unified hierarchical association framework}

To formalize non-linear extensions while maintaining clinical interpretability, we separate the functional shape of the scaling effect from its overall contribution to the log-hazard. We assume that the non-linear association effect $f(\nu)$ can be expressed as the product of the shared component and a scaling function $g(\nu)$:
\begin{equation}\label{eq:fg_decomp}
f(\nu) = g(\nu)\nu.
\end{equation}
By defining the non-linear effect as a multiplicative product, we fix the risk profile such that $f(0) = 0$, ensuring identifiability without arbitrarily restricting the functional shape. Because this anchors the reference baseline risk at $\nu=0$, we assume a value of zero is clinically meaningful and falls safely within the support of the observed data. Users should therefore center their longitudinal marker around a clinically meaningful reference value prior to model fitting. This reference does not need to be the sample mean or median, any value lying within the observed data support and carrying meaningful interpretation is appropriate \citep{lopez2025dealing}.

The possible shapes of the risk profile $f(\nu)$ depends entirely on the parameterization of the scaling function $g(\nu)$. Our goal is to achieve a clear hierarchy of nested models of varying complexity, where $f(\nu)$ is chosen to be strictly linear (Level 1), linear with a quadratic curvature (Level 2), or linear and quadratic with a smooth spline correction (Level 3). This ensures a natural nesting: if the spline correction is zero, Level 3 collapses to Level 2. If the quadratic term is zero, Level 2 collapses to Level 1.

We construct this hierarchy using a vector of basis coefficients $\bm{\gamma} = (\gamma_1, \dots, \gamma_K)^\top$. By making the choice to estimate one parameter, two parameters, or a larger set of parameters, the model transitions between these three structural levels:

\begin{enumerate}
    \item \textbf{Level 1 (Linear log-hazard):} With one parameter, $\bm{\gamma} = (\gamma_1)$, the choice for the scaling function is a constant, $g(\nu) = \gamma_1$, and the contribution to the log-hazard is strictly linear, $f(\nu) = \gamma_1 \nu$. This corresponds to the standard joint model assumption, where each unit increase in the biomarker produces the same multiplicative change in the hazard.
    \item \textbf{Level 2 (Quadratic log-hazard):} With two parameters, $\bm{\gamma} = (\gamma_1, \gamma_2)^\top$, the chosen scaling effect includes a linear trend across its domain, $g(\nu) = \gamma_1 + \gamma_2 \nu^s$, where $\nu^s$ represents an internally rescaled version of $\nu$, mapping it to a fixed interval (from $-0.5$ to $0.5$). Because $\nu^s$ is a linear transformation of $\nu$, their product naturally forms a second-degree polynomial. The contribution to the log-hazard then becomes quadratic, $f(\nu) = \gamma_1 \nu + \gamma_2 \nu^s \nu$. This accommodates global curvature such as U-shaped patterns.
    \item \textbf{Level 3 (Spline log-hazard):} With more parameters ($K > 2$), the remaining coefficients ($\gamma_3, \dots, \gamma_K$) define a data-driven smooth spline correction term $h(\nu)$, such that $g(\nu) = \gamma_1 + \gamma_2 \nu^s + h(\nu)$. The log-hazard flexibly accommodates localized departures from the parametric quadratic shape, $f(\nu) = \gamma_1 \nu + \gamma_2 \nu^s \nu + h(\nu)\nu$.
\end{enumerate}

While the observed biomarker is centered by the user to ensure $f(0) = 0$ represents a meaningful reference, the internal rescaling $\nu^s$ is performed to ensure numerical stability and scale-invariance when evaluating the splines.

To formulate the data-driven smooth spline correction $h(\nu)$ introduced in Level 3, the continuous scaling function $g(\nu)$ is evaluated at $K$ equidistant knots over the domain of the shared component. To ensure the resulting curve is smooth and prevents overfitting, we assign a second-order random walk (RW2) prior to these knot values. The corresponding RW2 precision matrix $\bm{Q}$ penalizes the squared second differences of the evaluated knots, encouraging smoothness.

A mathematical property of the RW2 prior is that it naturally separates into a simple parametric structure and flexible non-linear deviations. By applying an eigendecomposition to the scaled precision matrix, we obtain an orthogonal basis matrix $\bm{\Phi}$ with $K$ column vectors. A detailed derivation of this basis from the second-order random walk prior is provided in Section S1 of the Supplementary Materials. The first two basis vectors represent the unpenalized space, capturing a constant baseline $\bm{\phi}_1 = (1, \dots, 1)^\top$ and a standardized linear trend across the knots. The remaining basis vectors capture the data-driven non-linear deviations.

The values of the continuous scaling function $g(\nu)$ evaluated at the $K$ equidistant knots are defined as a linear combination of these basis vectors weighted by the coefficients $\bm{\gamma}$. The function is then evaluated at any arbitrary point $\nu$ via natural cubic spline interpolation of these nodal values. In this unified formulation, $\gamma_1$ controls the constant baseline scaling (Level 1), $\gamma_2$ controls the standardized linear trend of the scaling function across the domain (Level 2) and $\gamma_3, \dots, \gamma_K$ serve as deviation parameters for the localized adaptations $h(\nu)$ (Level 3).

In our experience, specifying $K=5$ knots (providing 3 deviation parameters) offers sufficient flexibility to adapt to complex biological curves without unnecessary parameterization. In this paper, we use $K=5$ knots for all the Level 3 models fitted.

\subsection{Bayesian inference and priors}

The proposed joint model is formulated as a latent Gaussian model (LGM). In an LGM, the observed data $\mathcal{D}_N$ are assumed conditionally independent given a latent field $\bm{u}$ and a vector of hyperparameters $\bm{\theta}$. The latent field $\bm{u}$ contains all unobserved model parameters, including the longitudinal fixed effects $\bm{\beta}$, the subject-specific random effects $\bm{b}_i$ and the survival fixed effects $\bm{\varphi}$. Hyperparameters include likelihood parameters such as the Gaussian residual error variance, random effects covariance terms and the precision parameter of the second-order random walk used to flexibly approximate the log-baseline hazard (see \cite{martino2011approximate} for more details). The association parameters $\bm{\gamma}$ are also estimated as hyperparameters within the INLA framework. A core requirement of the LGM framework is that $\bm{u}$ follows a multivariate Gaussian prior distribution characterized by a sparse precision matrix $\bm{Q}(\bm{\theta})$ (see Section S2 of the Supplementary Materials for details on how INLA deals with the survival likelihood).

We perform approximate Bayesian inference using the INLA algorithm \citep{rue2009approximate, vanniekerk2023new}. INLA computes highly accurate deterministic approximations of the posterior marginal distributions for the elements of $\bm{u}$ and $\bm{\theta}$ by relying on the sparsity of the precision matrix. This avoids the convergence issues and long computation times commonly associated with MCMC sampling in joint models.

Because the true trajectory $\nu_i(t)$ is an unobserved latent variable, the functional domain required to appropriately place the $K$ equidistant knots cannot be determined directly from the raw observed measurements. Moreover, because the non-linear scaling function $g(\nu)$ depends on this unobserved latent field, directly including the product $g(\nu)\nu$ violates the requirement that the linear predictor must be a linear combination of the latent field components. To resolve both issues and preserve the Latent Gaussian Model framework, our approach relies on an internal calibration to pre-compute the evaluation points for the non-linear scaling function.

First, a preliminary joint model assuming a standard linear association (Level 1) is fitted to obtain the posterior expectations of the shared latent field, denoted $\tilde{\nu} = \mathbb{E}[\nu \mid \mathcal{D}_N]$. The empirical minimum and maximum of these posterior estimates establish the functional domain over which the $K$ knots are placed equidistantly. Next, the non-linear scaling function $g()$ is evaluated at these pre-estimated points, providing the deterministic weight vector $g(\tilde{\nu})$. The term entering the survival linear predictor for the final model is thus $g(\tilde{\nu})\nu$. By treating $g(\tilde{\nu})$ as fixed evaluation weights, this product remains a strict linear combination of the latent field $\nu$. Of note, $\nu$ remains an active, jointly estimated shared component in this product. Because the model is fully joint, the longitudinal measurement error and uncertainty are appropriately propagated into the survival submodel. This formulation ensures the precision matrix remains sparse and the INLA algorithm can be applied without numerical approximations.

The prior specification for the basis coefficients $\bm{\gamma}$ acts as a regularization mechanism. The parametric components $\gamma_1$ (constant scaling) and $\gamma_2$ (linear scaling trend) are assigned weakly informative Gaussian priors, $\mathcal{N}(\text{mean}=0, \text{variance}=100)$, allowing the global shape of the association to be driven by the data likelihood. In contrast, estimating complex non-linear structures is computationally challenging and sensitive to data sparsity. For this reason, we propose conservative estimates for the non-linear deviations. The deviation parameters $\gamma_3, \dots, \gamma_K$ are assigned a zero-centered Laplace prior with a rate of 30. In Bayesian inference, this concentrated prior acts analogously to $L_1$ regularization (Lasso regression). It provides a strong shrinkage towards zero, slightly shifting the Bayesian posterior toward the conservative parsimonious model (Level 1 or Level 2) unless overcome by data evidence.

Model complexity and the plausibility of the linearity assumption can then be assessed using information criteria such as the Deviance Information Criterion (DIC) \citep{spiegelhalter2002bayesian} and Watanabe-Akaike Information Criterion (WAIC) \citep{watanabe2010asymptotic}, where lower values indicate a model that better captures the underlying data structure. The difference is a point estimate whose magnitude can only be interpreted relative to its sampling variability, the same value of $\Delta$WAIC may be decisive in a large cohort and inconclusive in a smaller one. On a single dataset, this variability can be estimated from the pointwise WAIC contributions, for two competing models, the observation-level differences in pointwise WAIC contributions gives both their mean (i.e., $\Delta$WAIC) and an empirical standard error, from which a standardized test statistic for the null hypothesis of equal predictive performance is derived \citep{vehtari2017practical}. The test therefore complements $\Delta$WAIC by indicating whether the observed difference is large relative to the noise, where a non-significant result means the two models are predictively indistinguishable given the available data, while a significant result indicates a systematic improvement. Because this procedure combines a Bayesian predictive criterion with a frequentist testing framework, the resulting p-values should be interpreted as complementary evidence rather than as a formal Bayesian decision rule. This pairwise comparison is implemented in \texttt{INLAjoint}. A summary of all model parameters and their assigned prior distributions is provided in Table S1 (Section S3 of the Supplementary Materials).

\section{Simulation study} \label{sec:simulations}

We designed a simulation study to evaluate the frequentist properties and robustness of the hierarchical framework. Specifically, we aimed to assess whether the proposed models accurately recover true non-linear associations and whether they avoid overfitting and inflating uncertainty when the true association is simpler.

\subsection{Simulation design}
We considered three scenarios characterized by different true underlying association structures. For each scenario, we generated 300 independent datasets, each including $N = 2000$ individuals followed up for a maximum of 10 time units. The longitudinal marker was simulated from a linear mixed-effects model with a fixed intercept, a linear time slope, a binary covariate and a time-by-covariate interaction, along with correlated individual random intercepts and slopes. This generated an average of 15 repeated measurements per individual. The detailed mathematical formulations of the models used for the simulation study are given in Section S4 of the Supplementary Materials.

Survival times were generated via a permutation algorithm \citep{sylvestre2008comparison} to appropriately align with the time-varying nature of the true unobserved longitudinal trajectory, with an average event rate of approximately 37\% across simulations. For each generated dataset, we fitted the three nested models based on the current value of the linear predictor:
\begin{enumerate}
    \item \textbf{Level 1 (Linear)}: Assumes a constant scaling effect, producing a linear effect on the log-hazard.
    \item \textbf{Level 2 (Quadratic)}: Incorporates a linear scaling trend, producing linear and quadratic effects on the log-hazard.
    \item \textbf{Level 3 (Spline)}: Incorporates the orthogonal basis (using $K=5$ knots), producing a smooth non-linear effect on the log-hazard.
\end{enumerate}
We evaluate parameter bias, standard deviation of the bias and frequentist 95\% coverage probabilities. Model fit was compared using $\Delta$DIC and $\Delta$WAIC, where the true data-generating model serves as the reference. For each replicate and each pair of models, we also applied the pairwise WAIC test introduced in Section~\ref{sec:methods} and report the proportion of the 300 replicates for which the test did not detect a statistically meaningful predictive difference between the two models at the conventional two-sided threshold $p > 0.05$. Of note, this proportion is a practical decision frequency on a per-dataset basis, not a nominal Type I error rate.

\subsection{Scenarios and results}

\begin{figure}[htpb]
    \centering
    \includegraphics[width=\textwidth]{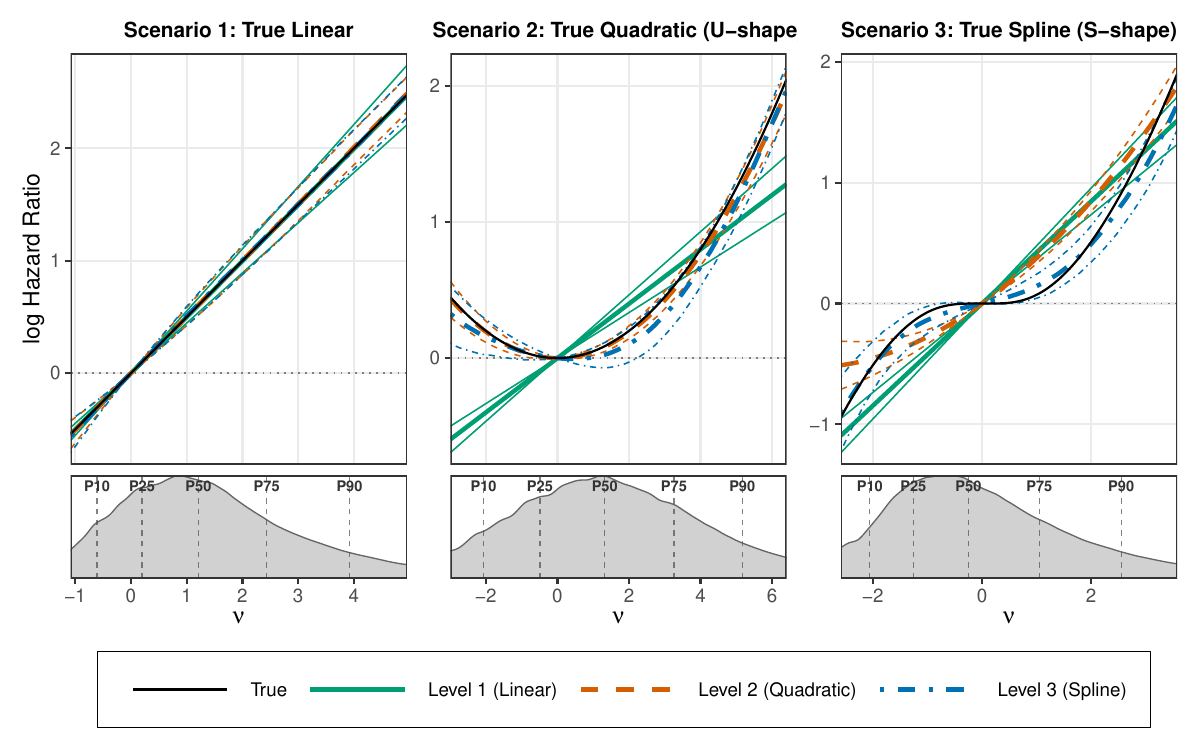}
    \caption{Estimated log hazard ratios across the three simulation scenarios comparing the Level 1 Linear (solid green), Level 2 Quadratic (dashed orange) and Level 3 Spline (dash-dotted blue) association models against the true underlying data generating mechanism (black solid line). Left: Scenario 1 (True Linear log-hazard). Middle: Scenario 2 (True Quadratic log-hazard). Right: Scenario 3 (True Spline log-hazard). The density plots below each plot display the distribution of the shared component $\nu$ in the simulated cohorts, with dashed vertical lines labeled P10, P25, P50, P75 and P90, corresponding to the 10th, 25th, 50th, 75th and 90th percentiles of this distribution.}
    \label{fig:simulations}
\end{figure}

\textbf{Scenario 1: True Linear log-hazard.}
The first scenario assumed a linear effect of the marker on the log-hazard ($f(\nu) = 0.5\nu$). The objective was to verify the robustness of the highly parameterized quadratic and flexible models when no non-linear flexibility is required.

As shown in Table \ref{tab:sim_linear}, all three models performed identically in recovering the true fixed and random effects. The Level 2 and Level 3 models successfully shrank their additional parameters towards zero, perfectly aligning their estimated curves with the true linear trajectory (Figure \ref{fig:simulations}, left panel). Coverage probabilities remained near their nominal 95\% levels across all parameters. The mean $\Delta$DIC and $\Delta$WAIC relative to the true Level 1 model were small and marginally in favor of the more flexible specifications (within 3.5 units on average, with a standard deviation of a similar order of magnitude across replicates). This small residual advantage reflects the fact that the additional parameters of the more flexible specifications can absorb residual noise, which translates into a marginal apparent improvement in fit. DIC and WAIC have a tendency to insufficiently penalize unnecessary parameters under heavy shrinkage \citep{vehtari2017practical}. Because the mean differences are of the same order of magnitude as their across-replicate standard deviation, they cannot be separated from sampling variability, a diagnosis confirmed by the pairwise WAIC test applied within each replicate. The test classified the Level 2 and Level 3 models as predictively indistinguishable from the true Level 1 model in 82.7\% and 81.3\% of the replicates, reflecting the penalty for unnecessary parameters.

\begin{table}[H]
\centering
\small
\caption{Simulation results for Scenario 1: True Linear log-hazard (n=2000, nsim=300). Results are presented as: bias (SD of bias) [coverage \%].}
\begin{tabular}{lcrrr}
\toprule
Parameter & True value & Level 1 (Linear) & Level 2 (Quadratic) & Level 3 (Spline)\\
\midrule
\multicolumn{5}{l}{\textit{Fixed effects}}\\
$\beta_0$ & 0.00 & -0.003 (0.028) [92\%] & -0.003 (0.028) [92\%] & -0.003 (0.028) [92\%]\\
$\beta_1$ & 0.30 & 0.001 (0.011) [95\%] & 0.001 (0.011) [95\%] & 0.001 (0.011) [95\%]\\
$\beta_2$ & 0.50 & 0.002 (0.039) [93\%] & 0.002 (0.039) [93\%] & 0.002 (0.039) [94\%]\\
$\beta_3$ & 0.20 & 0.000 (0.016) [96\%] & 0.000 (0.016) [96\%] & 0.000 (0.016) [96\%]\\
$\sigma_\varepsilon$ & 0.30 & 0.000 (0.001) [94\%] & 0.000 (0.001) [94\%] & 0.000 (0.001) [95\%]\\
\midrule
\multicolumn{5}{l}{\textit{Random effects}}\\
$\sigma_{b_0}$ & 0.80 & -0.001 (0.013) [96\%] & -0.001 (0.013) [96\%] & -0.001 (0.013) [95\%]\\
$\sigma_{b_1}$ & 0.30 & 0.000 (0.006) [96\%] & 0.000 (0.006) [97\%] & 0.000 (0.006) [96\%]\\
$\rho_{b_0 b_1}$ & 0.30 & -0.001 (0.026) [94\%] & -0.002 (0.027) [95\%] & -0.001 (0.026) [90\%]\\
\midrule
\multicolumn{5}{l}{\textit{Association}}\\
$\gamma_1$ & 0.50 & -0.003 (0.026) [96\%] &  & \\
\midrule
\multicolumn{5}{l}{\textit{Model comparison}}\\
\multicolumn{2}{l}{$\Delta$DIC} & Reference & $-2.4$ (2.5) & $-3.4$ (3.4)\\
\multicolumn{2}{l}{$\Delta$WAIC} & Reference & $-2.3$ (1.8) & $-3.1$ (2.4)\\
\multicolumn{2}{l}{Pairwise WAIC test (\% non-sig.)} & Reference & 82.7\% & 81.3\%\\
\multicolumn{2}{l}{CPU time in sec.} & 35.3 (4.4) & 74.7 (6.9) & 101.3 (15.4)\\
\bottomrule
\end{tabular}
\label{tab:sim_linear}
\end{table}

\textbf{Scenario 2: True Quadratic log-hazard (U-shape).}
Motivated by non-monotonic relationships observed in epidemiology, the true association in this scenario followed a purely quadratic log-hazard function: $f(\nu) = \nu^2 / 20$.

The results (Table \ref{tab:sim_quadratic}) demonstrated that the standard Level 1 model failed to capture the relationship, producing highly misspecified fits due to attempting to average the opposing directions of risk across the biomarker's range (Figure \ref{fig:simulations}, middle panel). Both the Level 2 and Level 3 models accurately captured the true U-shaped curve. Because the true relationship was perfectly quadratic, the Level 3 model effectively shrank its additional RW2 components to zero. The information criteria heavily penalized the misspecified Level 1 model (mean $\Delta$DIC above 200 units) and the pairwise WAIC test rejected Level 1 in favor of Level 2 in 100\% of the replicates. The Level 2 and Level 3 models produced nearly indistinguishable fits, with the raw mean $\Delta$DIC and $\Delta$WAIC showing a small residual preference of about 5.5 units for the more flexible Level 3 specification, again accompanied by an across-replicate standard deviation of the same order of magnitude. This preference reflects the same mechanisms described in Scenario 1. The pairwise WAIC test applied this penalty and reported the two models as predictively indistinguishable in 86.7\% of the replicates, identifying the parsimonious true Level 2 model as the best structure.

\begin{table}[H]
\centering
\small
\caption{Simulation results for Scenario 2: True Quadratic log-hazard (n=2000, nsim=300). Results are presented as: bias (SD of bias) [coverage \%]. True association: $f(\nu) = \nu^2/20$ (U-shaped).}
\begin{tabular}{lcrrr}
\toprule
Parameter & True value & Level 1 (Linear) & Level 2 (Quadratic) & Level 3 (Spline)\\
\midrule
\multicolumn{5}{l}{\textit{Fixed effects}}\\
$\beta_0$ & 0.00 & -0.006 (0.069) [93\%] & -0.006 (0.069) [92\%] & -0.006 (0.069) [92\%]\\
$\beta_1$ & 0.30 & 0.002 (0.015) [95\%] & 0.000 (0.015) [95\%] & 0.000 (0.015) [95\%]\\
$\beta_2$ & 0.50 & 0.005 (0.097) [94\%] & 0.005 (0.097) [94\%] & 0.005 (0.097) [94\%]\\
$\beta_3$ & 0.20 & -0.002 (0.021) [94\%] & -0.001 (0.021) [95\%] & -0.002 (0.021) [95\%]\\
$\sigma_\varepsilon$ & 0.30 & 0.000 (0.001) [93\%] & 0.000 (0.001) [92\%] & 0.000 (0.001) [93\%]\\
\midrule
\multicolumn{5}{l}{\textit{Random effects}}\\
$\sigma_{b_0}$ & 2.00 & -0.001 (0.031) [97\%] & -0.001 (0.031) [96\%] & -0.002 (0.031) [96\%]\\
$\sigma_{b_1}$ & 0.40 & -0.002 (0.008) [95\%] & -0.001 (0.008) [95\%] & -0.001 (0.008) [95\%]\\
$\rho_{b_0 b_1}$ & 0.20 & -0.007 (0.027) [92\%] & -0.003 (0.027) [94\%] & -0.005 (0.027) [94\%]\\
\midrule
\multicolumn{5}{l}{\textit{Model comparison}}\\
\multicolumn{2}{l}{$\Delta$DIC} & 208.1 (29.5) & Reference & $-5.6$ (4.6)\\
\multicolumn{2}{l}{$\Delta$WAIC} & 210.2 (29.5) & Reference & $-5.5$ (4.2)\\
\multicolumn{2}{l}{Pairwise WAIC test (\% non-sig.)} & 0.0\% & Reference & 86.7\%\\
\multicolumn{2}{l}{CPU time in sec.} & 30.1 (3.4) & 64.5 (4.4) & 88.3 (7.5)\\
\bottomrule
\end{tabular}
\label{tab:sim_quadratic}
\end{table}

\textbf{Scenario 3: True Spline log-hazard (S-shape).}
The third scenario introduced a complex, S-shaped risk trajectory: $f(\nu) = \nu \log(\nu^2 + 1)^{3/2} / 8$. The log hazard ratio has a sharp increase across negative predictor values, a plateau near the zero reference and increases again across positive values. Such shapes mimic complex dynamics increasingly encountered in health data \citep{prezelin2023longitudinal}.

Under this non-linearity, the Level 3 model adapted to the complex curve, accurately capturing the steep risk gradients at both extremes as well as the plateau (Figure \ref{fig:simulations}, right panel). The Level 2 model provided a rough global approximation ($\Delta$DIC = 47.5 compared to Level 3), lacking the localized flexibility to capture the risk plateau. While Level 1 performed poorly ($\Delta$DIC = 87.5), its poor fit was less pronounced than observed in Scenario 2. This occurs because a straight line can roughly approximate the average trajectory of an S-shape, whereas it completely misses a U-shape. As a result, the Level 3 model produced the best fit among the three approaches, verifying the need for this level of flexibility (Table \ref{tab:sim_spline}). The pairwise WAIC test rejected both Level 1 and Level 2 in favor of Level 3 in 100\% of the replicates, confirming on a per-dataset basis that the additional flexibility captures a structural signal rather than residual noise. To assess the robustness of the framework with a smaller sample size, we also evaluated this scenario using datasets of $N=500$ individuals (results detailed in Table S2 and Figure S1, Section S5 of the Supplementary Materials). The smaller sample exhibited similar patterns, with increased attenuation bias at the boundaries due to reduced data density in the extremes. The $\Delta$DIC and $\Delta$WAIC remained positive, confirming the Spline model as preferred, although with a reduced sample size the pairwise WAIC test has less statistical power to resolve smaller differences between nested models, so the non-significant rates reported in Table~\ref{tab:sim_spline_n500} for Level 2 against Level 3 increase accordingly while the Level 1 model is still rejected in more than 90\% of the replicates.

\begin{table}[H]
\centering
\small
\caption{Simulation results for Scenario 3: True Spline log-hazard (n=2000, nsim=300). Results are presented as: bias (SD of bias) [coverage \%]. True association: $f(\nu) = \nu \cdot \log(\nu^2+1)^{3/2}/8$ (S-shaped).}
\begin{tabular}{lcrrr}
\toprule
Parameter & True value & Level 1 (Linear) & Level 2 (Quadratic) & Level 3 (Spline)\\
\midrule
\multicolumn{5}{l}{\textit{Fixed effects}}\\
$\beta_0$ & -1.50 & -0.002 (0.028) [93\%] & -0.002 (0.028) [93\%] & -0.002 (0.028) [92\%]\\
$\beta_1$ & 0.30 & 0.001 (0.011) [95\%] & 0.000 (0.011) [95\%] & -0.001 (0.011) [95\%]\\
$\beta_2$ & 0.50 & 0.002 (0.039) [94\%] & 0.002 (0.039) [94\%] & 0.002 (0.039) [94\%]\\
$\beta_3$ & 0.20 & 0.001 (0.016) [94\%] & 0.001 (0.016) [94\%] & 0.001 (0.016) [95\%]\\
$\sigma_\varepsilon$ & 0.30 & 0.000 (0.001) [94\%] & 0.000 (0.001) [94\%] & 0.000 (0.001) [94\%]\\
\midrule
\multicolumn{5}{l}{\textit{Random effects}}\\
$\sigma_{b_0}$ & 0.80 & 0.000 (0.013) [95\%] & 0.000 (0.013) [95\%] & 0.000 (0.013) [96\%]\\
$\sigma_{b_1}$ & 0.30 & 0.001 (0.006) [95\%] & 0.001 (0.006) [94\%] & 0.001 (0.006) [95\%]\\
$\rho_{b_0 b_1}$ & 0.30 & 0.000 (0.030) [90\%] & 0.000 (0.027) [92\%] & -0.002 (0.028) [91\%]\\
\midrule
\multicolumn{5}{l}{\textit{Model comparison}}\\
\multicolumn{2}{l}{$\Delta$DIC} & 87.5 (16.3) & 47.5 (12.7) & Reference\\
\multicolumn{2}{l}{$\Delta$WAIC} & 89.4 (16.2) & 47.1 (12.4) & Reference\\
\multicolumn{2}{l}{Pairwise WAIC test (\% non-sig.)} & 0.0\% & 0.0\% & Reference\\
\multicolumn{2}{l}{CPU time in sec.} & 28.5 (3.6) & 60.6 (5.4) & 84.4 (14.7)\\
\bottomrule
\end{tabular}
\label{tab:sim_spline}
\end{table}

Across all scenarios, coverage probabilities remained robust. The framework demonstrated high computational efficiency. Fitting the full Level 3 model on datasets of 2000 individuals required an average of less than 2 minutes on a server (with 16 CPU cores and 45 GB RAM).

\subsection{Pointwise evaluation and structural properties}

Across all scenarios, frequentist coverage probabilities for the global structural parameters remained highly robust, staying near their nominal 95\% levels (Tables 1 to 3). To characterize the behavior of the flexible association models, we evaluated the absolute bias and frequentist coverage of the estimated log-hazard function $f(\nu)$ at the 10th, 25th, 50th, 75th and 90th percentiles of the true $\nu$ distribution (Table S3, Section S5 of the Supplementary Materials).

This pointwise evaluation highlights standard properties of penalized spline estimation \citep{ruppert2003semiparametric}. Within the data-dense interquartile range, the flexible models accurately recover the non-linear trajectories with negligible bias. The main exception occurs at the median of the true $\nu$ distribution in the two Spline scenarios, where coverage is reduced despite the correct model. This median lies near $\nu = 0$, where $f(0) = 0$ by construction. Around this central point, the true curve passes through a short flat region between two increasing segments, and the smooth spline fit slightly rounds this flat region. The resulting estimate is extremely precise, so even a very small residual bias is sufficient to reduce frequentist coverage.

At the unobserved extremes, the framework intentionally exhibits a conservative attenuation bias, visibly flattening the estimated curves. This boundary behavior is driven by two factors. First, pre-computing the evaluation points introduces a regression dilution effect. Because $\tilde{\nu}$ extracted from the preliminary model are naturally shrunk toward the population mean, evaluating the non-linear scaling function at these coordinates slightly attenuates the evaluated weights at the boundaries. The model is fully joint, so the estimation uncertainty of the active latent field $\nu$ is propagated, but the structural shape itself is evaluated at these slightly shrunk points.

Second, this behavior is reinforced by the conservative Laplace priors, which systematically shrink the non-linear deviations toward zero in sparse data regions. Given the highly precise standard errors generated by the large sample size, this conservative shift leads to lower frequentist coverage of the credible intervals at the boundaries. This behavior is further illustrated by the supplementary $N=500$ Spline scenario (Table S2, Section S5 of the Supplementary Materials). With a smaller sample size, the reduced data density in the extremes amplifies the impact of the informative Laplace priors, leading to an increasing attenuation bias toward the null in these sparse regions.

This conservative boundary attenuation is clinically desirable, as it prevents the flexible splines from generating unstable risk estimates in regions with sparse data. Ultimately, the framework reliably recovers the complex non-linear dynamics and overall clinical risk shapes where the vast majority of patient data resides.

\section{Application: Joint modeling of body mass index dynamics and all-cause mortality} \label{sec:application}

To illustrate the practical value of the proposed methodology, we applied our framework to characterize the complex association between longitudinal trajectories of body mass index (BMI) and the risk of all-cause mortality.

\subsection{Epidemiological context and clinical motivation}
The relationship between BMI and mortality is widely documented in the literature, consistently demonstrating a non-linear, U-shaped risk curve. In adults, both underweight and obesity confer elevated mortality risks \citep{di2016body, zheng2013obesity}. However, focusing only on the value of BMI masks a central temporal dimension: the rate of change (slope) of an individual's weight over time. Unintentional weight loss in older adults is often a highly prognostic indicator of underlying frailty or undiagnosed disease. Trajectory analyses consistently demonstrate that this weight loss carries a higher mortality risk than BMI's value itself \citep{zajacova2014body}. On the other hand, individuals who maintain a stable weight or transition from normal to overweight status in later adulthood often exhibit the lowest long-term mortality risks \citep{zheng2021life}.

Addressing this epidemiological question requires the simultaneous assessment of both the Current Value (CV) and Current Slope (CS) of the biomarker, while adequately accounting for measurement error and informative dropout. This joint parameterization offers a unique epidemiological advantage by enabling the unbiased evaluation of both components simultaneously: it isolates the effect of the slope for a given value and independently evaluates the risk of the value for a given slope. Standard joint models are conceptually well suited for this task but have imposed a linear assumption on both components. This constraint can mask threshold dynamics or U-shaped biological realities, leading to biased clinical conclusions. Evaluating the non-linear effects of multiple time-dependent shared components simultaneously has been methodologically limited due to significant computational constraints. Our unified INLA-based framework permits the simultaneous inclusion of multiple non-linear association structures to achieve this evaluation.

\subsection{Study data and model specification}

We analyzed publicly available data from the Health and Retirement Study (HRS), a nationally representative longitudinal panel study of individuals over age 50 in the United States \citep{HRS_Public_Data, RAND_HRS_File}. The dataset contains repeated biennial measurements of self-reported BMI and ascertained times of death through linkage to the National Death Index. Out of a total sample of 45,234 individuals, our analysis included 42,403 individuals who had complete baseline covariates and at least one BMI measurement, providing a total of 287,225 repeated BMI measurements.

Table \ref{tab:app_desc_full} summarizes the baseline characteristics of the sample. Median follow-up from study entry was 24.3 years, and 18,917 participants (44.6\%) died. Participants contributed a median of 6 BMI measurements (range: 1 to 16). Additional data descriptions, including individual BMI trajectories and Kaplan-Meier survival estimates, are provided in Figure S2 (Section S6 of the Supplementary Materials).

\begin{table}[htpb]
\centering
\caption{Baseline characteristics of the Health and Retirement Study ($N=42,403$).}
\begin{tabular}{lrlr}
\toprule
\textbf{Demographics} & \textbf{Value (\%)} & \textbf{Clinical Characteristics} & \textbf{Value (\%)} \\
\midrule
\textit{Age at baseline} & & \textit{Baseline Comorbidities} & \\
\quad $<55$ years (Reference) & 42.7 & \quad Smoking history (Ever) & 57.5 \\
\quad $55 - 60$ years & 21.9 & \quad Hypertension & 39.4 \\
\quad $>60$ years & 35.4 & \quad Diabetes & 13.1 \\
Gender (Female) & 56.0 & \quad Heart disease & 14.2 \\
\textit{Race} & & \quad Stroke & 4.8 \\
\quad White (Reference) & 72.3 & \quad Cancer & 7.1 \\
\quad Black & 19.4 & \textit{Baseline BMI (kg/m$^2$)} & \\
\quad Other & 8.3 & \quad\quad Mean (SD) / Median & 27.7 (5.9) / 27.0 \\
& & \quad\quad Underweight ($<$18.5) & 1.7 \\
& & \quad\quad Normal (18.5--25) & 31.9 \\
& & \quad\quad Overweight (25--30) & 38.2 \\
& & \quad\quad Obese ($\geq$30) & 28.2 \\
\bottomrule
\end{tabular}
\label{tab:app_desc_full}
\end{table}

To correct for right-skewness and heteroskedasticity, we modeled a log-transformed version of BMI, centered around 27 kg/m$^2$: $y_{ij} = \log(\text{BMI}_{ij}/27)$. This centering value corresponds to the median BMI across all repeated observations in the sample and was identified as the value minimizing the risk of death in preliminary analyses. Time $t$ was rescaled such that one unit corresponds to a decade. The longitudinal submodel included a fixed intercept and a linear time slope. To account for established demographic differences in aging trajectories, the submodel included interactions between time and both age categories and gender. This structure acknowledges that weight changes over time naturally differ between men and women and across different baseline age groups. We also included individual-level random intercepts and random slopes. The model was further adjusted for race, smoking status and five baseline binary comorbidities: hypertension, diabetes, heart disease, stroke and cancer.

The survival submodel for all-cause mortality was also adjusted for baseline age category, gender, race, smoking history, hypertension, diabetes, heart disease, stroke and cancer. The baseline hazard was approximated flexibly using a second-order random walk with 15 knots. A summary of the specified priors is provided in Table S1 (Section S3 of the Supplementary Materials). The estimated baseline hazard curve is provided in Figure S3 (Section S6 of the Supplementary Materials).

We investigated the joint impact of two components of the transformed BMI trajectory simultaneously:
\begin{enumerate}
\item \textbf{Current Value (CV)}: The expected level of the transformed BMI at time $t$.
\item \textbf{Current Slope (CS)}: The expected instantaneous rate of change (first derivative) of the transformed BMI at time $t$. Because we modeled the natural logarithm of BMI, the mathematical derivative ($\frac{d}{dt}\ln(X) = \frac{X'}{X}$) represents the instantaneous relative rate of change of the marker.
\end{enumerate}

We fitted three joint models, incrementally increasing the flexibility of the association structure for both the CV and CS components simultaneously: Level 1 (Linear), Level 2 (Quadratic) and Level 3 (Spline). Of note, while the framework does not theoretically restrict the CV and CS components to share the same structural level, we chose to increase flexibility for both components simultaneously because the higher levels naturally nest the lower ones, allowing each component to adapt to its own degree of complexity within the same model. The detailed mathematical formulations of the models used for the application are given in Section S4 of the Supplementary Materials.

\subsection{Results}

\begin{figure}[htpb]
\centering
\includegraphics[width=\textwidth]{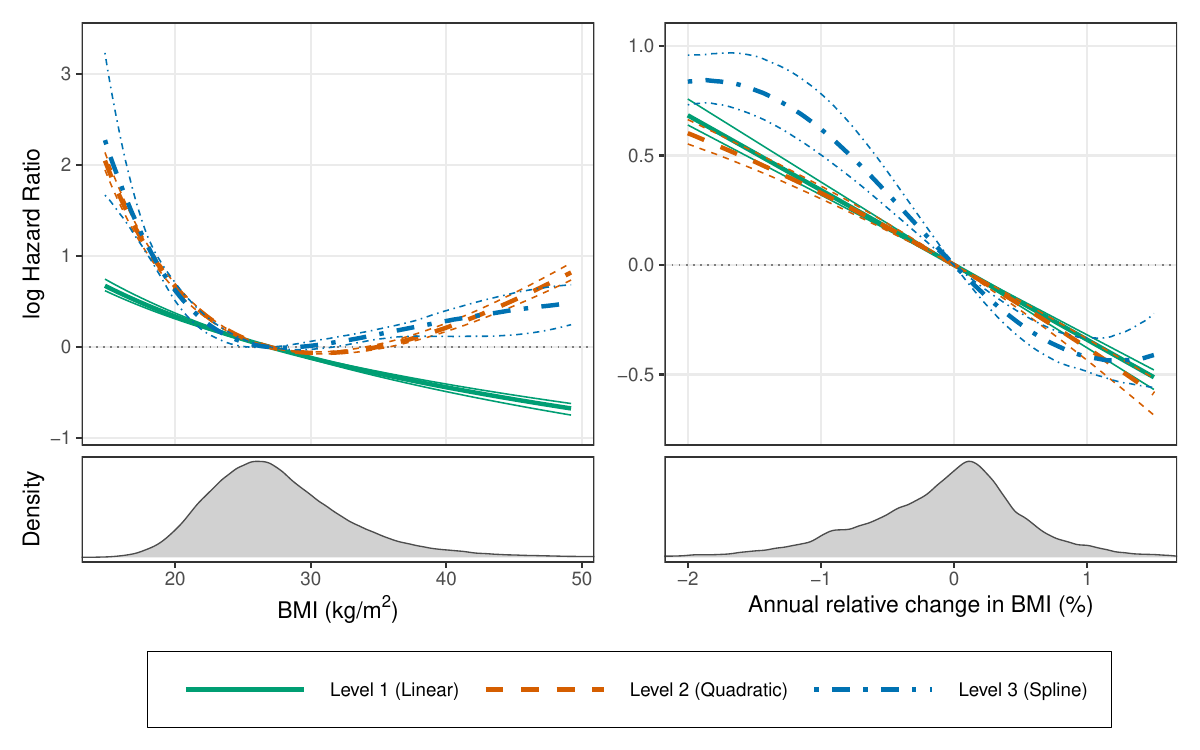}
\caption{Estimated log-hazard ratios for all-cause mortality associated with the Current Value (left) and Current Slope (right) of BMI from the full HRS multi-cohort, adjusted for baseline age category, gender, race, smoking history, hypertension, diabetes, heart disease, stroke and cancer (see Table \ref{tab:app_results_full}). Density plots below the curves indicate the empirical distribution of the latent shared components. Of note, because the longitudinal submodel is fitted to log-transformed BMI, the linear effect (Level 1) natively exhibits a slight curvature when plotted against the original absolute BMI scale (kg/m$^2$) on the x-axis. For the current slope, the derivative of log-BMI corresponds to the relative rate of change per decade, which was scaled to reflect the annual relative change in BMI (\%).}
\label{fig:application_results}
\end{figure}

Plotting the estimated log-hazard ratios conditional on the shared latent predictors showed some non-linear relationships for the two components (Figure \ref{fig:application_results}). To ensure rigorous clinical interpretation, marginal density plots of the latent shared processes are displayed below the hazard curves, showing where the data provide information to estimate the risk robustly.

For the Current Value (CV) of BMI (adjusted for the current rate of change), the non-linear models identified a U-shaped risk profile. The Level 3 Spline model demonstrated that mortality risk is minimized near a BMI of $27$ kg/m$^2$, a finding confirmed by a separate analysis using only the current value association without the slope component (results not shown). Deviations from this nadir carry an increased risk. For instance, an underweight patient with a BMI of $20$ kg/m$^2$ faces an adjusted hazard ratio (HR) of 1.86 [95\% credible interval: 1.64, 2.04] compared to the reference, while extreme obesity (BMI of $40$ kg/m$^2$) elevates mortality risk to an HR of 1.32 [1.10, 1.50]. Examining the marginal density plot below the curves, the majority of observations lie between 20 and 40 kg/m$^2$, aligned with the narrowest credible intervals. In the tails, the uncertainty bands for the Spline model widen, reflecting an uncertainty related to the density of the data rather than constrained by parametric assumptions. The linear Level 1 model failed to identify this dynamic, averaging the opposing directions of risk to estimate an HR of 0.64 at BMI 40. While the Quadratic and Spline models agreed on the increased risk for low BMI, a difference emerged for overweight and moderately obese patients (BMI 30 to 35 kg/m$^2$). The Quadratic model estimates a nadir beyond BMI 30, resulting in hazard ratios below 1 in this range. In contrast, the Spline model estimates a slightly higher log-hazard ratio, indicating an elevated risk (HR $> 1$) throughout the 30 to 40 kg/m$^2$ range.

In evaluating the Current Slope (CS) effect (adjusted for the current BMI value), the results confirmed that weight loss carries an increased mortality risk in this population. Specifically, the Spline model estimates that an annual relative decrease of 1\% in body mass is associated with a higher hazard of death (HR = 1.87 [1.64, 2.29]) compared to a stable weight trajectory (0\% change). A moderate weight gain (+1\% per year) exhibits a protective, lower hazard (HR = 0.66 [0.61, 0.73]). To provide clinical context, a 1\% relative increase for an individual with a baseline BMI of 25 kg/m$^2$ corresponds to an absolute increase of 0.25 BMI points per year, or approximately 0.7 kg. While the Level 1 model restricts this dynamic to a linear trend, the Level 3 Spline model uncovers a non-linear S-shape, capturing plateau effects for both low and high slopes. The Spline model estimates a higher hazard ratio for weight loss compared to the Quadratic model (1.87 [1.64, 2.29] versus 1.39 [1.35, 1.44] at -1\%, relative to a stable weight trajectory at 0\% change) and a similar hazard ratio for weight gain (0.66 versus 0.68 at +1\%). The credible intervals widen at the extremes where the density of observed slopes is minimal, although uncertainty is quite low overall given that we have a large sample size.

The information criteria confirmed that relaxing the linearity assumption improved the model fit. The WAIC and DIC penalized the linear approach and favored the most flexible Level 3 Spline model ($\Delta$DIC = 1902.9 compared to Linear and $\Delta$DIC = 790.8 compared to Quadratic, see Table \ref{tab:app_results_full}). This aligns with the visual evidence and our simulations. While the CV effect is globally quadratic, the Spline model allows for deviations from the parametric trajectory, which accounts for the statistical difference observed in the 30 to 35 kg/m$^2$ range. Moreover, the flexibility of the splines is required to capture the plateauing deviations observed in the Current Slope, supporting the added complexity of this model. To corroborate these results, a formal model comparison test based on observation-level WAIC differences confirmed that the Level 3 model consistently fits the observed data better than both Level 1 ($p < 0.001$) and Level 2 ($p < 0.001$), and the Level 2 model over Level 1 ($p < 0.001$). Table \ref{tab:app_results_full} presents the complete structure of the estimated parameters for the survival and longitudinal submodels.

\begin{table}[htpb]
\centering
\small
\renewcommand{\arraystretch}{0.95}
\caption{Estimated parameters from the Level 3 (Spline) joint model applied to the full HRS multi-cohort sample. Survival submodel effects are reported as hazard ratios (HR). Model comparison criteria (DIC, WAIC) are provided for all three levels.}
\begin{tabular}{lcc}
\toprule
\textbf{Parameter} & \textbf{Mean} & \textbf{95\% Credible interval} \\
\midrule
\multicolumn{3}{l}{\textit{Survival submodel (Hazard Ratios)}} \\
Age: $55 - 60$ years & 1.368 & [1.310, 1.428] \\
Age: $>60$ years & 3.419 & [3.310, 3.531] \\
Gender (Female) & 0.731 & [0.711, 0.751] \\
Race (Black) & 1.051 & [1.009, 1.094] \\
Race (Other) & 0.823 & [0.764, 0.886] \\
Smoking (Ever) & 1.242 & [1.209, 1.277] \\
Hypertension & 1.223 & [1.186, 1.261] \\
Diabetes & 1.710 & [1.639, 1.784] \\
Heart disease & 1.571 & [1.512, 1.632] \\
Stroke & 1.733 & [1.633, 1.840] \\
Cancer & 1.467 & [1.394, 1.543] \\
\midrule
\multicolumn{3}{l}{\textit{Longitudinal submodel (Log-BMI centered at 27)}} \\
Intercept & 0.017 & [0.013, 0.022] \\
Time (decades) & 0.007 & [0.005, 0.009] \\
Age: $55 - 60$ years & -0.020 & [-0.024, -0.015] \\
Age: $>60$ years & -0.087 & [-0.091, -0.082] \\
Gender (Female) & -0.013 & [-0.017, -0.009] \\
Race (Black) & 0.044 & [0.039, 0.048] \\
Race (Other) & 0.011 & [0.005, 0.018] \\
Smoking (Ever) & -0.016 & [-0.020, -0.013] \\
Hypertension & 0.075 & [0.071, 0.078] \\
Diabetes & 0.086 & [0.081, 0.092] \\
Heart disease & -0.006 & [-0.011, -0.001] \\
Stroke & -0.032 & [-0.041, -0.024] \\
Cancer & -0.009 & [-0.016, -0.002] \\
Time $\cdot$ Age: $55 - 60$ & -0.015 & [-0.017, -0.012] \\
Time $\cdot$ Age: $>60$ & -0.069 & [-0.071, -0.066] \\
Time $\cdot$ Gender (Female) & 0.002 & [-0.001, 0.004] \\
\midrule
\multicolumn{3}{l}{\textit{Random components}} \\
$\sigma_{b_0}$ (Random intercept) & 0.181 & [0.179, 0.182] \\
$\sigma_{b_1}$ (Random slope) & 0.081 & [0.080, 0.083] \\
$\rho$ (Correlation) & -0.106 & [-0.133, -0.074] \\
$\sigma_e$ (Residual error) & 0.069 & [0.069, 0.069] \\
\midrule
\multicolumn{3}{l}{\textit{Model comparison}} \\
$\Delta$DIC vs Level 3 (Level 1 / Level 2) & \multicolumn{2}{c}{1902.9 / 790.8} \\
$\Delta$WAIC vs Level 3 (Level 1 / Level 2) & \multicolumn{2}{c}{2062.6 / 1055.5} \\
Pairwise WAIC test ($z$, $p$-value) & & \\
\quad Level 1 vs Level 2 & \multicolumn{2}{c}{$z = 5.77$, $p < 0.001$} \\
\quad Level 1 vs Level 3 & \multicolumn{2}{c}{$z = 11.27$, $p < 0.001$} \\
\quad Level 2 vs Level 3 & \multicolumn{2}{c}{$z = 7.69$, $p < 0.001$} \\
CPU time in sec. (Level 1 / Level 2 / Level 3) & \multicolumn{2}{c}{556 / 1646 / 1761} \\
\bottomrule
\end{tabular}
\label{tab:app_results_full}
\end{table}

As an exploratory follow-up motivated by the possibility that the effect of weight change depends on the current BMI level, we fitted separate stratified models within the normal-weight, overweight and obese subgroups (results not shown). All three strata produced qualitatively similar non-linear patterns, with the Level 3 model preferred in each case. This suggests that BMI categories may be too imprecise to reveal a continuous interaction between current BMI and weight change, or that the excess mortality risk associated with obesity is largely captured by the current value component, while the current slope reflects a more general dynamic signal related to frailty or underlying disease. Under this interpretation, weight loss remains a marker of vulnerability and moderate weight gain of relative stability across BMI strata.

\section{Discussion} \label{sec:discussion}

The assumption of a linear association between longitudinal biomarkers and the risk of clinical events is a common simplification in joint modeling \citep{wulfsohn1997joint, rizopoulos2012joint}. The limitations of the linearity assumption have been widely discussed for standard covariates, such as the effect of age on survival. The shared longitudinal marker is in the exact same situation when scaled by a single constant. Extending the joint modeling framework to allow for non-linearity in the marker's effect is a natural and necessary improvement to capture the true underlying risk. While mathematically convenient, the linear assumption is systematically adopted without formal verification. This is largely due to a lack of accessible and computationally efficient diagnostic tools. Our proposed hierarchical framework addresses this gap, allowing researchers to estimate non-linear association structures efficiently.

By using the INLA algorithm, we formulated the non-linear association applying an orthogonal basis expansion governed by a second-order random walk. The application of a Laplace prior serves as an automated regularization mechanism, pulling data-driven deviations toward zero and safely reducing the model to a parsimonious linear or quadratic structure on the log-hazard unless contradicted by the data. This ensures that in smaller samples, the absence of evidence naturally shrinks the non-linear deviations toward zero, safely collapsing the model into a stable, parsimonious parametric structure rather than overfitting noise. Moreover, the scale of this prior can be easily adjusted by the user to be more or less informative depending on the specific clinical context. This unified hierarchy allows researchers to incrementally increase flexibility and rely on information criteria like WAIC and DIC, complemented by one-to-one model comparison tests based on observation-level WAIC differences, to evaluate the plausibility of the linearity assumption.

Relaxing the linearity assumption for covariate effects in survival models is well established in the standard Cox model, where penalized splines are routinely used to model non-linear covariate-hazard relationships, for example via the \texttt{pspline()} term in the \texttt{coxph} function from the \texttt{survival} package \citep{therneau2000cox}. Our framework extends this principle to the more complex setting of shared latent processes in joint models. While previous efforts to introduce joint models with non-linear associations provided excellent theoretical foundations \citep{kohler2018nonlinear}, our approach successfully overcomes the computational bottlenecks and convergence challenges associated with sampling-based estimation. The computation times recorded in our application requiring less than 30 minutes (with 24 CPU cores and 90 GB RAM) to fit the most complex Spline joint model for a cohort of over 42,000 subjects and 287,000 total repeated measurements illustrate the unprecedented scalability of this framework.

The application to the HRS dataset highlights the unique capabilities of the method. We simultaneously estimated non-linear structures for two distinct shared components, the current value and the current slope of BMI. The framework correctly identified a U-shaped risk curve for BMI's current value while uncovering a complex S-shaped risk trajectory for relative weight change. This proves the capacity of the method to isolate and differentiate complex functional forms within the same patient cohort, evaluating the slope conditional on the value and the value conditional on the slope, without requiring arbitrary categorizations prior to analysis.

A methodological limitation of our framework is the internal calibration of the non-linear evaluation points. To preserve the computational advantages of the latent Gaussian model structure natively within INLA, we evaluate the non-linear scaling weights $g(\tilde{\nu})$ at pre-computed posterior expectations of the latent field. While the shared component $\nu$ remains fully jointly estimated to ensure that longitudinal measurement error and uncertainty are correctly propagated into the survival submodel, the evaluation points themselves are subject to natural shrinkage toward the population mean. As demonstrated in our pointwise simulation evaluations, evaluating the function at these shrunk coordinates induces a systematic conservative attenuation bias. While this pre-evaluation step introduces regression dilution, the framework remains superior to the two-stage approach because it fully propagates longitudinal uncertainty through the latent field $\nu$. The extent of the dilution bias depends on the true underlying shape. It peaks for purely quadratic associations where the preliminary linear model fails to capture the relationship and adjust for informative censoring, but remains smaller for shapes like S-curves that can be roughly approximated by a linear trend. Moreover, our use of informative Laplace priors for numerical stability shifts the estimates toward the conservative model in regions lacking data. As a result, while the joint model accurately recovers the global non-linear trajectory, it conservatively flattens extreme risk estimates toward the null and leads to an expected drop in frequentist coverage at the tails of the biomarker distribution.

We argue that this structurally conservative boundary effect is a methodological strength. By forcing the model to be conservative, we avoid hallucinating non-linear effects and overfitting noise. Along with massive gains in computational speed, this allows researchers to routinely test the linearity assumption in modern cohorts without risking unstable spline extrapolations. To ensure rigorous clinical interpretation, we advise plotting the marginal density of the shared latent process alongside the estimated hazard curves, as demonstrated in our application, to restrict clinical interpretations to data-dense regions.

Future extensions of this framework could include alternative shared components, such as the cumulative area under the curve (AUC), which are completely compatible with the proposed structural parameterization. The framework could also be extended to multiple longitudinal markers with non-linear association with one or multiple event times (e.g., recurrent events, competing risks, etc.). As precision medicine increasingly depends on characterizing individualized dynamic risk profiles rather than population-average trends, the ability to reliably estimate these complex, multi-dimensional non-linear trajectories with high computation speed represents a significant advancement in biostatistical methodology. An important clinical question is whether the effect of weight change varies continuously with an individual's current BMI level. Addressing this question would require extending the present framework to model an interaction between the current value and the current slope, which remains an important direction for future work. To share the method and facilitate its application, we provide a reproducible R script based on one dataset from our simulation studies (Section S7 of the Supplementary Materials).  Moreover, both the INLA algorithm and the \texttt{INLAjoint} R package used for our analyses are free and open-source software, making this advanced framework readily accessible to the biostatistical community.

\section*{Acknowledgements}
The HRS (Health and Retirement Study) is sponsored by the National Institute on Aging (grant numbers NIA U01AG009740 and NIA R01AG073289) and is conducted by the University of Michigan.

\section*{Conflict of Interest}
The authors declare no potential conflict of interests.

\section*{Data Availability Statement}
The data from the Health and Retirement Study (HRS) used in Section \ref{sec:application} are publicly available for reasearchers from the University of Michigan (\url{https://hrsdata.isr.umich.edu/}). The \texttt{INLAjoint} R package is freely available on CRAN. Reproducible code to demonstrate the proposed methodology is provided in the Supplementary Materials.

\printbibliography

\clearpage
\section*{Supplementary Materials} \label{sec:supp}

\renewcommand{\thetable}{S\arabic{table}}
\setcounter{table}{0}
\renewcommand{\thefigure}{S\arabic{figure}}
\setcounter{figure}{0}

\subsection*{S1. Mathematical derivation of the orthogonal basis from the second-order random walk}

This section details the mathematical construction of the orthogonal basis matrix $\bm{\Phi}$ used to model the non-linear scaling function $g(\nu)$ introduced in Section \ref{sec:methods}.

Let $\bm{g} = (g_1, \dots, g_K)^\top$ denote the vector of the continuous scaling function evaluated at $K$ equidistant knots across the domain of the shared component. To ensure the resulting interpolated curve is smooth, we assign a second-order random walk (RW2) prior to these knot values. The RW2 prior penalizes the squared second differences between adjacent knots, defined as $\Delta^2 g_k = (g_k - g_{k-1}) - (g_{k-1} - g_{k-2}) = g_k - 2g_{k-1} + g_{k-2}$.

Within the latent Gaussian model framework, the joint prior distribution for the vector $\bm{g}$ is a multivariate normal distribution $\mathcal{N}(\bm{0}, \bm{Q}^{-1})$, where $\bm{Q}$ is the sparse precision matrix encoding these second differences. The mathematical penalty added to the model log-likelihood is directly proportional to the quadratic form $\bm{g}^\top \bm{Q} \bm{g}$.

An essential mathematical property of the RW2 penalty is its behavior when the sequence of knots forms a perfectly flat line or a linear trend. For these specific shapes, all second differences evaluate to exactly zero. As a result, the penalty $\bm{g}^\top \bm{Q} \bm{g}$ evaluates to zero. In linear algebra, the space of vectors that result in a zero penalty forms the null space of the matrix $\bm{Q}$. This means the RW2 prior does not penalize constant baselines or linear trends. It only penalizes deviations from a straight line.

To explicitly decompose the unpenalized parametric shapes from the flexible data-driven deviations, we compute the eigendecomposition of the precision matrix:
\begin{equation}
 \bm{Q} = \bm{E} \bm{\Omega} \bm{E}^\top
\end{equation}
where $\bm{\Omega}$ is a diagonal matrix of eigenvalues $\omega_1, \dots, \omega_K$ and $\bm{E}$ is a matrix of corresponding orthogonal column eigenvectors $\bm{e}_1, \dots, \bm{e}_K$.

Because the null space of $\bm{Q}$ has a rank of two, the first two eigenvalues are exactly zero ($\omega_1 = 0$, $\omega_2 = 0$). The first eigenvector $\bm{e}_1$ represents a constant flat line $(1, \dots, 1)^\top$. The second eigenvector $\bm{e}_2$ represents a standardized linear trend across the knots. The remaining eigenvalues ($\omega_3, \dots, \omega_K$) are strictly positive. Their corresponding eigenvectors represent non-linear curves of increasing complexity.

If we parameterized the splines directly using these raw eigenvectors, the non-linear deviation parameters would face different penalty strengths. The penalty applied to each deviation parameter would be proportional to its specific eigenvalue $\omega_k$. To standardize this penalty across all possible deviations, we scale the non-linear eigenvectors by the inverse square root of their respective positive eigenvalues. This defines the final orthogonal basis matrix $\bm{\Phi}$ with $K$ column vectors:
\begin{align}
\bm{\phi}_1 &= \bm{e}_1 \nonumber \\
\bm{\phi}_2 &= \bm{e}_2 \nonumber \\
\bm{\phi}_k &= \frac{\bm{e}_k}{\sqrt{\omega_k}} \quad \text{for } k = 3, \dots, K
\end{align}

We then express the vector of evaluated knots as a linear combination of this scaled basis weighted by the coefficients $\bm{\gamma} = (\gamma_1, \dots, \gamma_K)^\top$, such that $\bm{g} = \bm{\Phi} \bm{\gamma}$.

The mathematical advantage of this transformation becomes clear when we substitute $\bm{g} = \bm{\Phi} \bm{\gamma}$ into the original penalty term:
\begin{align}
    \bm{g}^\top \bm{Q} \bm{g} &= (\bm{\Phi} \bm{\gamma})^\top \bm{Q} (\bm{\Phi} \bm{\gamma}) \nonumber \\
    &= \bm{\gamma}^\top (\bm{\Phi}^\top \bm{Q} \bm{\Phi}) \bm{\gamma} \nonumber \\
    &= \sum_{k=3}^K \gamma_k^2
\end{align}

By scaling the basis vectors, the eigenvalues cancel out. The complex matrix formulation of the variance penalty simplifies to a standardized sum of squares. Every non-linear shape is penalized on an equal scale.

The parameter $\gamma_1$ controls the unpenalized constant baseline $\bm{\phi}_1$ and $\gamma_2$ controls the unpenalized linear trend $\bm{\phi}_2$. We assign a zero-centered Laplace prior to the remaining deviation parameters $\gamma_3, \dots, \gamma_K$. Because the non-linear basis vectors are perfectly standardized, this concentrated prior acts as an automated regularization mechanism. It forces unnecessary non-linear flexibility to zero unless the data provide sufficient evidence for a more complex shape.

\subsection*{S2. Data expansion and the Poisson approximation for the survival likelihood}
A primary computational bottleneck in joint modeling is the evaluation of the survival likelihood when the hazard function includes time-varying endogenous covariates. Because the cumulative hazard requires integrating these evolving trajectories over time, standard algorithms rely on time-consuming numerical approximations such as Gauss quadrature.

The INLA methodology implemented in the \texttt{INLAjoint} package avoids this issue by using the Poisson approximation of the Cox proportional hazards model \citep{holford1976life, whitehead1980fitting}. The continuous follow-up time is discretized into a fine grid of small intervals. Within each interval, the hazard is assumed to be approximately constant, allowing the continuous-time survival process to be modeled as a sequence of discrete-time event counts following a Poisson distribution.

By restructuring the data so that each subject contributes a separate row for each time interval they are at risk, the evolving time-dependent shared components can be evaluated specifically for each interval. In addition to handling time-varying covariates, this data expansion allows the baseline hazard to be modeled flexibly using a random walk over the discrete intervals. This decomposition converts the complex time-dependent integral of the joint likelihood into a standard latent Gaussian model structure \citep{martino2011approximate}. As a result, INLA processes the joint model using highly efficient sparse matrix algebra, avoiding the need for numerical quadrature and reducing computation time.

\subsection*{S3. Summary of model parameters and prior distributions}
To ensure transparency and reproducibility, Table \ref{tab:supp_priors} summarizes the prior distributions assigned to the parameters within the \texttt{INLAjoint} implementation. Of note, these are default specifications that can be modified by the user, as the INLA framework easily accommodates alternative prior distributions.

\begin{table}[H]
\centering
\small
\caption{Summary of model parameters and prior distributions.}
\label{tab:supp_priors}
\begin{tabular}{cll}
\toprule
\textbf{Symbol} & \textbf{Description} & \textbf{Prior Distribution} \\
\midrule
\multicolumn{3}{l}{\textit{Latent Field Components ($\bm{u}$)}} \\
$\bm{\beta}$ & Longitudinal fixed effects & $\mathcal{N}(\text{mean}=0, \text{variance} = 100)$ \\
$\bm{\varphi}$ & Survival fixed effects & $\mathcal{N}(\text{mean}=0, \text{variance} = 100)$ \\
$\bm{b}_i$ & Subject-specific random effects & $\mathcal{N}(\text{mean} = \bm{0}, \text{covariance} = \bm{\Sigma}_b)$ \\
\midrule
\multicolumn{3}{l}{\textit{Hyperparameters ($\bm{\theta}$)}} \\
$\sigma_e^2$ & Residual error variance & Inverse-Gamma (shape = 1, scale = $5 \cdot 10^{-5}$) \\
$\bm{\Sigma}_b$ & Random effects covariance & Inverse-Wishart (df = 3, scale matrix = $\bm{I})$ \\
$\sigma_{\tau}$ & Baseline hazard RW2 standard deviation & PC prior: $\text{Prob}(\sigma_{\tau} > 0.5) = 0.01$ \\
\midrule
\multicolumn{3}{l}{\textit{Association Parameters ($\bm{\gamma}$)}} \\
$\gamma_1$ & Association: Constant baseline & $\mathcal{N}(\text{mean}=0, \text{variance} = 100)$ \\
$\gamma_2$ & Association: Linear trend & $\mathcal{N}(\text{mean}=0, \text{variance} = 100)$ \\
$\gamma_3 \dots \gamma_K$ & Association: Spline deviations & $\mathcal{L}\textit{aplace}(\text{location}=0, \text{rate} = 30)$ \\
\bottomrule
\end{tabular}
\end{table}

\subsection*{S4. Details on the models}

This section details the equations of the joint models evaluated in the simulation study and the application.

\subsubsection*{Simulation study models}

The true unobserved trajectory $\eta_i(t)$ includes a fixed intercept ($\beta_0$), a linear time slope ($\beta_1$), a binary baseline covariate ($X_i$, with effect $\beta_2$) and a time-by-covariate interaction ($\beta_3$), along with subject-specific random intercepts ($b_{0i}$) and random slopes ($b_{1i}$):
\begin{equation}
    \eta_i(t) = (\beta_0 + b_{0i}) + (\beta_1 + b_{1i})t + \beta_2 X_i + \beta_3 (X_i \cdot t)
\end{equation}
The subject-specific random effects follow a bivariate normal distribution, $\bm{b}_i = (b_{0i}, b_{1i})^\top \sim \mathcal{N}(\bm{0}, \bm{\Sigma}_b)$.

For the survival submodel, because we simulate data using a permutation algorithm \citep{sylvestre2008comparison}, the baseline hazard $\lambda_0(t)$ is unspecified. To focus the evaluation purely on the association structure, no other survival covariates are included. The hazard function depends on the longitudinal process through the current value of the marker ($\nu_i(t) = \eta_i(t)$):
\begin{equation}
    \lambda_i(t) = \lambda_0(t) \exp\left( f(\eta_i(t)) \right)
\end{equation}
The baseline hazard is approximated by a second-order random walk with 15 knots. The specific structural shape of $f()$ (Linear, Quadratic, or Spline) varies according to the three data-generating scenarios.

\subsubsection*{Application models}

For the application to the Health and Retirement Study, the longitudinal outcome is the log-transformed and centered Body Mass Index, $y_{ij} = \log(\text{BMI}_{ij}/27)$. The true unobserved trajectory $\eta_i(t)$ is modeled as:
\begin{equation}
    \eta_i(t) = (\beta_0 + b_{0i}) + \bm{X}_{1i}^\top \bm{\beta}_1 + \left( \beta_t + b_{1i} + \bm{X}_{2i}^\top \bm{\beta}_2 \right) t
\end{equation}
where $\bm{X}_{1i}$ represents the vector of baseline covariates adjusting the intercept (age categories, gender, race, smoking status, hypertension, diabetes, heart disease, stroke and cancer) and $\bm{X}_{2i}$ represents the vector of demographic covariates interacting with time (age categories and gender).

The survival submodel for all-cause mortality evaluates the joint non-linear effect of both the current value ($\nu_{1i}(t) = \eta_i(t)$) and the current slope ($\nu_{2i}(t) = \frac{\partial}{\partial t}\eta_i(t)$) of the biomarker simultaneously:
\begin{equation}
    \lambda_i(t) = \lambda_0(t) \exp\left( \bm{w}_i^\top \bm{\varphi} + f_{CV}(\eta_i(t)) + f_{CS}\left(\frac{\partial}{\partial t}\eta_i(t)\right) \right)
\end{equation}
where $\lambda_0(t)$ is the baseline hazard approximated by a second-order random walk with 15 knots, $\bm{w}_i$ is the vector of baseline survival covariates (age categories, gender, race, smoking status and the five clinical comorbidities) and $\bm{\varphi}$ represents their corresponding fixed effects. The non-linear scaling functions $f_{CV}()$ and $f_{CS}()$ evaluate the respective risk profiles for the BMI value and its instantaneous relative rate of change across the three hierarchical flexibility levels.

\subsection*{S5. Additional simulation results}

\begin{figure}[H]
\centering
\includegraphics[width=\textwidth]{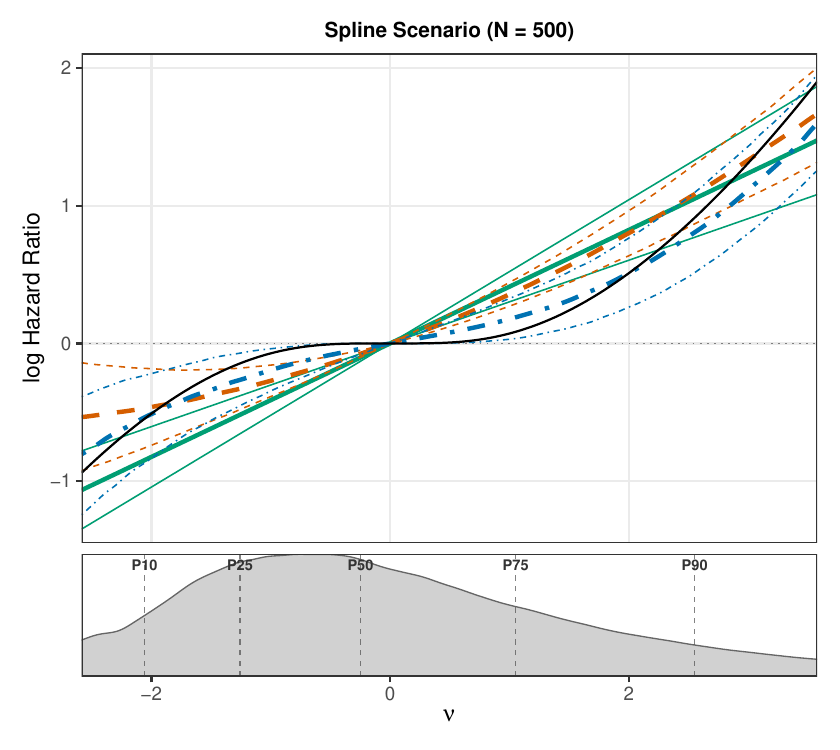}
\caption{Estimated log Hazard Ratio for the Spline Scenario evaluated on a smaller sample size of $N=500$. The plot compares the estimated association curves from the Level 1 (Linear), Level 2 (Quadratic) and Level 3 (Spline) joint models against the true underlying data-generating S-shaped mechanism (black line) across the domain of the shared latent component $\nu$.}
\label{fig:supp_data_desc_spline500}
\end{figure}

\begin{table}[H]
\centering
\small
\caption{Simulation results for the Spline log-hazard scenario evaluated on a smaller sample size ($N=500$, nsim=300). Results are presented as: bias (empirical standard deviation) [coverage \%]. True association: $f(\nu) = \nu \cdot \log(\nu^2+1)^{3/2}/8$ (S-shaped).}
\label{tab:sim_spline_n500}
\begin{tabular}{lcrrr}
\toprule
Parameter & True value & Level 1 (Linear) & Level 2 (Quadratic) & Level 3 (Spline)\\
\midrule
\multicolumn{5}{l}{\textit{Fixed effects}}\\
$\beta_0$ & -1.50 & -0.004 (0.051) [94\%] & -0.004 (0.051) [94\%] & -0.004 (0.051) [94\%]\\
$\beta_1$ & 0.30 & 0.002 (0.023) [94\%] & 0.001 (0.023) [93\%] & 0.000 (0.023) [93\%]\\
$\beta_2$ & 0.50 & -0.001 (0.073) [94\%] & -0.001 (0.073) [94\%] & -0.001 (0.073) [94\%]\\
$\beta_3$ & 0.20 & -0.001 (0.032) [94\%] & -0.001 (0.032) [94\%] & -0.002 (0.032) [94\%]\\
$\sigma_\varepsilon$ & 0.30 & 0.000 (0.003) [95\%] & 0.000 (0.003) [95\%] & 0.000 (0.003) [95\%]\\
\midrule
\multicolumn{5}{l}{\textit{Random effects}}\\
$\sigma_{b_0}$ & 0.80 & 0.000 (0.026) [97\%] & 0.000 (0.026) [97\%] & 0.000 (0.026) [97\%]\\
$\sigma_{b_1}$ & 0.30 & 0.005 (0.012) [92\%] & 0.005 (0.012) [92\%] & 0.004 (0.012) [92\%]\\
$\rho_{b_0 b_1}$ & 0.30 & -0.003 (0.048) [97\%] & -0.001 (0.048) [95\%] & -0.005 (0.048) [85\%]\\
\midrule
\multicolumn{5}{l}{\textit{Model comparison}}\\
\multicolumn{2}{l}{$\Delta$DIC} & 22.0 (9.0) & 11.1 (6.3) & Reference\\
\multicolumn{2}{l}{$\Delta$WAIC} & 22.6 (9.1) & 11.1 (6.2) & Reference\\
\multicolumn{2}{l}{Pairwise WAIC test (\% non-sig.)} & 7.3\% & 34.7\% & Reference\\
\multicolumn{2}{l}{CPU time in sec.} & 9.4 (1.3) & 19.1 (2.0) & 23.5 (3.8)\\
\bottomrule
\end{tabular}
\end{table}

\begin{table}[H]
\centering
\small
\caption{Pointwise statistics for $f(\nu)$ evaluated at P10, P25, P50, P75 and P90, corresponding to the 10th, 25th, 50th, 75th and 90th percentiles of the true $\nu$ distribution, across all simulation scenarios. Results are presented as: bias (SD of bias) [coverage \%]. Models correspond to Level 1 (Linear), Level 2 (Quadratic) and Level 3 (Spline).}
\label{tab:pw_unscaled_stats_supp}
\begin{tabular}{llrrrr}
\toprule
Scenario & Percentile & Level 1 & Level 2 & Level 3 \\
\midrule
\multirow{5}{*}{Linear} & P10 & 0.013 (0.084) [96\%] & -0.025 (0.268) [94\%] & -0.088 (0.467) [91\%] \\
 & P25 & 0.016 (0.052) [94\%] & -0.005 (0.143) [94\%] & -0.038 (0.228) [91\%] \\
 & P50 & -0.020 (0.001) [0\%] & -0.020 (0.002) [0\%] & -0.021 (0.003) [0\%] \\
 & P75 & -0.016 (0.052) [94\%] & -0.005 (0.076) [92\%] & -0.004 (0.075) [92\%] \\
 & P90 & -0.013 (0.084) [96\%] & 0.000 (0.098) [87\%] & -0.009 (0.097) [88\%] \\
\midrule
\multirow{5}{*}{Quadratic} & P10 & -2.227 (0.092) [0\%] & -0.158 (0.132) [73\%] & -0.314 (0.176) [51\%] \\
 & P25 & -1.091 (0.057) [0\%] & -0.066 (0.068) [81\%] & -0.177 (0.096) [63\%] \\
 & P50 & -0.012 (0.001) [0\%] & 0.000 (0.001) [91\%] & 0.002 (0.002) [84\%] \\
 & P75 & 0.101 (0.057) [49\%] & -0.076 (0.041) [38\%] & -0.191 (0.081) [25\%] \\
 & P90 & -0.307 (0.092) [6\%] & -0.174 (0.072) [15\%] & -0.237 (0.093) [15\%] \\
\midrule
\multirow{5}{*}{Spline} & P10 & 0.157 (0.076) [60\%] & 0.957 (0.116) [0\%] & -0.050 (0.361) [89\%] \\
 & P25 & -0.326 (0.047) [0\%] & 0.054 (0.060) [93\%] & 0.010 (0.105) [95\%] \\
 & P50 & -0.017 (0.001) [0\%] & -0.013 (0.001) [0\%] & -0.004 (0.003) [59\%] \\
 & P75 & 0.326 (0.047) [0\%] & 0.326 (0.048) [1\%] & -0.031 (0.085) [91\%] \\
 & P90 & -0.157 (0.076) [60\%] & 0.031 (0.088) [92\%] & -0.206 (0.124) [52\%] \\
\midrule
\multirow{5}{*}{Spline ($N=500$)} & P10 & 0.192 (0.163) [84\%] & 0.925 (0.249) [10\%] & 0.295 (0.362) [76\%] \\
 & P25 & -0.304 (0.101) [17\%] & 0.049 (0.126) [96\%] & -0.003 (0.163) [95\%] \\
 & P50 & -0.017 (0.002) [0\%] & -0.013 (0.002) [1\%] & -0.006 (0.004) [52\%] \\
 & P75 & 0.304 (0.101) [17\%] & 0.282 (0.110) [13\%] & -0.005 (0.165) [85\%] \\
 & P90 & -0.192 (0.163) [84\%] & -0.066 (0.213) [91\%] & -0.230 (0.214) [64\%] \\
\bottomrule
\end{tabular}
\end{table}

\subsection*{S6. Application data descriptions and baseline hazard}

\begin{figure}[H]
\centering
\includegraphics[width=\textwidth]{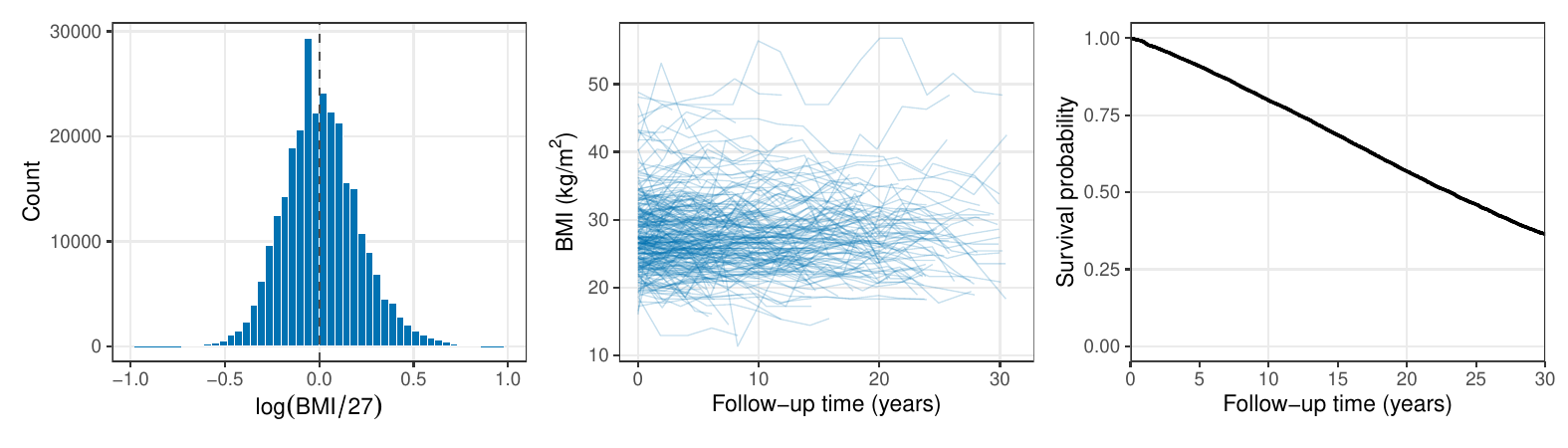}
\caption{Descriptive plots for the Health and Retirement Study (HRS) data. (Left) Histogram of $\log(\mathrm{BMI}/27)$, the centered longitudinal outcome used in the joint model, with the dashed line indicating the reference value $\log(27/27)=0$. (Middle) Spaghetti plot illustrating longitudinal BMI trajectories for a random subsample of individuals. (Right) Kaplan-Meier estimate of overall survival probability, with a 95\% pointwise confidence band that is not visible due to the large sample size.}
\label{fig:supp_data_desc_hrs}
\end{figure}

\begin{figure}[H]
\centering
\includegraphics[width=0.5\textwidth]{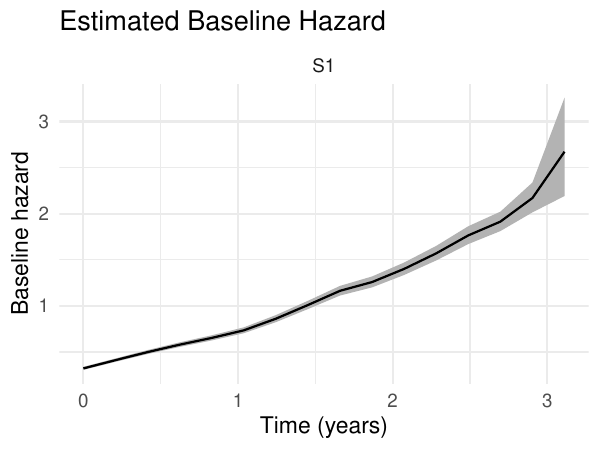}
\caption{Estimated baseline hazard (Level 3 Model) for all-cause mortality in the Health and Retirement Study (HRS) application. The plot illustrates the baseline hazard over time (decades), which was approximated using a second-order random walk with 15 knots, with the shaded region indicating the corresponding 95\% credible interval.}
\label{fig:supp_baseline_hazard}
\end{figure}

\subsection*{S7. Reproducible code and software implementation}
The proposed methodology is fully implemented within the free open-source \texttt{INLAjoint} R package. To facilitate dissemination and allow researchers to easily adopt the framework, we provide an open-source code script based on a single generated dataset from Scenario 3 of our simulation studies. This reproducible code is publicly available on GitHub at \texttt{\url{https://github.com/DenisRustand/JM_NL_code/}}. The provided code illustrates how to structure the data, define the joint models with the 3 levels of flexibility described in the paper, performs a formal model comparison test between nested models and plot the association curves.

\end{document}